\documentclass[twocolumn]{svjour3}          
\smartqed  
\usepackage{latexsym}
\usepackage{amssymb}
\usepackage{amsfonts}
\usepackage{amsmath}
\usepackage{bm}
\usepackage{graphicx}   
\usepackage{color}
\usepackage{subfigure}
\usepackage{times}
\usepackage{units}
\usepackage{hyperref}
\begin{document}

\title{Multi-Messenger Astrophysics with THESEUS in the 2030s}


\author{Riccardo~Ciolfi         \and
        Giulia~Stratta  \and 
      Marica~Branchesi \and 
       Bruce~Gendre \and 
           Stefan~Grimm  \and          
        Jan~Harms  \and          
        Gavin~Paul~Lamb  \and 
          Antonio~Martin-Carrillo  \and          
        Ayden~McCann  \and          
        Gor~Oganesyan  \and 
        Eliana~Palazzi  \and          
        Samuele~Ronchini  \and          
        Andrea~Rossi  \and 
        Om~Sharan~Salafia  \and 
      Lana~Salmon \and 
       Stefano~Ascenzi \and 
           Antonio~Capone  \and          
        Silvia~Celli  \and          
        Simone~Dall'Osso  \and 
         Irene~Di~Palma  \and          
        Michela~Fasano  \and          
        Paolo~Fermani  \and 
       Dafne~Guetta  \and          
       Lorraine~Hanlon  \and          
       Eric~Howell  \and 
       Stephane~Paltani  \and 
       Luciano~Rezzolla \and 
       Serena~Vinciguerra \and 
        Angela~Zegarelli  \and          
        Lorenzo~Amati  \and          
        Andrew~Blain  \and 
        Enrico~Bozzo  \and          
        Sylvain~Chaty  \and          
        Paolo~D'Avanzo  \and 
        Massimiliano~De~Pasquale  \and          
        H\"usne~Dereli-B\'egu\'e  \and          
        Giancarlo~Ghirlanda  \and 
        Andreja~Gomboc  \and 
       Diego~G\"otz \and 
       Istvan~Horvath \and
       Rene~Hudec \and 
        Luca~Izzo  \and          
        Emeric~Le~Floch  \and        
        Liang Li \and  
        Francesco~Longo  \and 
        S.~Komossa  \and          
        Albert~K.~H.~Kong  \and          
       Sandro~Mereghetti  \and 
       Roberto~Mignani  \and          
       Antonios~Nathanail  \and          
       Paul~T.~O'Brien  \and 
       Julian~P.~Osborne  \and 
      Asaf~Pe'er \and 
       Silvia~Piranomonte \and 
       Piero~Rosati  \and          
       Sandra~Savaglio  \and          
      Fabian~Sch\"ussler  \and          
       Olga~Sergijenko  \and 
        Lijing~Shao  \and          
       Nial~Tanvir  \and          
        Sara~Turriziani  \and 
       Yuji~Urata  \and          
        Maurice~van~Putten  \and          
        Susanna~Vergani  \and 
        Silvia~Zane  \and 
        Bing~Zhang   
}


\institute{Riccardo Ciolfi \at
              INAF, Osservatorio Astronomico di Padova, Vicolo dell'Osservatorio 5, I-35122 Padova, Italy;  
              INFN, Sezione di Padova, Via Francesco Marzolo 8, I-35131 Padova, Italy \\
              \email{riccardo.ciolfi@inaf.it}           
           \and
           Giulia Stratta \at
            INAF, Osservatorio di Astrofisica e Scienza dello Spazio,  via Piero Gobetti 93/3, 40129 Bologna, Italy; 
            INFN, Sezione di Firenze, via Sansone 1, I-50019, Firenze, Italy
           \and
	Marica Branchesi, Stefan Grimm, Jan Harms, Gor Oganesyan, Samuele Ronchini \at
	Gran Sasso Science Institute, Viale F. Crispi 7, I-67100 L'Aquila (AQ), Italy; 
	INFN, Laboratori Nazionali del Gran Sasso, I-67100 Assergi, Italy
           \and
	Bruce Gendre, Ayden McCann, Eric Howell \at
	OzGrav-UWA, University of Western Australia, 35 Stirling Highway, M013, 6009 Crawley, WA, Australia
           \and
	Gavin Paul Lamb, Andrew Blain, Paul T. O'Brien, Julian P. Osborne, Nial Tanvir \at
	School of Physics and Astronomy, University of Leicester, University Road, Leicester, LE1 7RH, UK
           \and
	Antonio Martin-Carrillo, Lana Salmon, Lorraine Hanlon \at
	School of Physics and Centre for Space Research, University College Dublin, Dublin 4, Ireland
           \and
	Eliana Palazzi, Andrea Rossi, Lorenzo Amati \at
	INAF, Osservatorio di Astrofisica e Scienza dello Spazio,  via Piero Gobetti 93/3, 40129 Bologna, Italy
           \and
	Om Sharan Salafia, Paolo D'Avanzo, Giancarlo Ghirlanda \at
	INAF, Osservatorio Astronomico di Brera, Via E. Bianchi 46, 23807 Merate, Italy; 
	INFN, Sezione di Milano-Bicocca, Piazza della Scienza 3, 20126 Milano, Italy
           \and
	Stefano Ascenzi \at
	Institute of Space Sciences (ICE, CSIC), Campus UAB, Carrer de Can Magrans s/n, 08193, Barcelona, Spain; 
	Institut d'Estudis Espacials de Catalunya (IEEC), Carrer Gran Capita 2-4, 08034 Barcelona, Spain
           \and
	Antonio Capone, Silvia Celli, Irene Di Palma, Michela Fasano, Paolo Fermani, Angela Zegarelli \at
	Dipartimento di Fisica dell'Universit\`a La Sapienza, P.le Aldo Moro 2, I-00185 Rome, Italy; 
	INFN, Sezione di Roma, P.le Aldo Moro 2, I-00185 Rome, Italy
           \and
	Simone Dall'Osso \at
	Gran Sasso Science Institute, Viale F. Crispi 7, I-67100 L'Aquila (AQ), Italy
           \and
	Dafne Guetta \at
	ORT-Braude College, Carmiel, Israel
           \and
	Stephane Paltani \at
	Department of Astronomy, University of Geneva, 1205 Versoix, Switzerland
           \and
	Luciano Rezzolla \at
	Institut f\"ur Theoretische Physik, Max-von-Laue-Strasse 1, D-60438 Frankfurt, Germany; 
	Frankfurt Institute for Advanced Studies, Ruth-Moufang-Strasse 1, D-60438 Frankfurt, Germany; 
	School of Mathematics, Trinity College, Dublin 2, Ireland
           \and
	Serena Vinciguerra \at
	Anton Pannekoek Institute for Astronomy, University of Amsterdam, Science Park 904, 1090GE Amsterdam, The Netherlands
           \and
	Enrico Bozzo \at
	Department of Astronomy, University of Geneva, Chemin d'Ecogia 16, CH-1290 Versoix, Switzerland
           \and
	Sylvain Chaty \at
	Universit\'e de Paris and Universit\'e Paris Saclay, CEA, CNRS, AIM, F-91190 Gif-sur-Yvette, France; 
	Universit\'e de Paris, CNRS, AstroParticule et Cosmologie, F-75013 Paris, France
           \and
	Massimiliano De Pasquale \at
	Department of Astronomy and Space Sciences, Istanbul University, Beyaz{\i}t 34119, Istanbul, Turkey
           \and
	H\"usne Dereli-B\'egu\'e \at
	Max Planck Institute for Extraterrestrial Physics, Giessenbachstrasse 1, D-85748 Garching, Germany; 
	Department of Physics, Bar-Ilan University, Ramat-Gan 52900, Israel
           \and
	Andreja Gomboc \at
	Center for Astrophysics and Cosmology, University of Nova Gorica, Vipavska 13, SI-5000 Nova Gorica, Slovenia
           \and
	Diego G\"otz \at
	AIM-CEA/DRF/Irfu/D\'epartement d'Astrophysique, CNRS, Universit\'e Paris-Saclay, Universit\'e de Paris, Orme des Merisiers, F-91191 Gif-sur-Yvette, France
           \and
           Istvan Horvath \at
           University of Public Service, Budapest, Hungary
           \and           
	Rene Hudec \at
	Czech Technical University in Prague, Faculty of Electrical Engineering, Prague, Czech Republic; 
	Astronomical Institute, Czech Academy of Sciences, Ondrejov, Czech Republic; 
	Kazan Federal University, Kazan, Russian Federation
           \and
	Luca Izzo \at
	DARK, Niels Bohr Institute, University of Copenhagen, Lyngbyvej 2, 2100 Copenhagen, Denmark
           \and
	Emeric Le Floch \at
	Laboratoire AIM, CEA/DSM/IRFU, CNRS, Universit\'e  Paris-Diderot, Bat. 709, 91191 Gif-sur-Yvette, France
           \and
            Liang Li \at
            ICRANet, Piazza della Repubblica 10, I-65122 Pescara, Italy
           \and
	Francesco Longo \at
	Universit\`a degli Studi di Trieste, via Valerio 2, I-34127 Trieste, Italy; 
	INFN, Sezione di Trieste, via Valerio 2, I-34127 Trieste, Italy; 
	Institute for Fundamental Physics of the Universe (IFPU), I-34151 Trieste, Italy
           \and
	S.~Komossa \at
	Max-Planck Institut f\"ur Radioastronomie, Auf dem H\"ugel 69, 53111 Bonn, Germany
           \and
	Albert K. H. Kong \at
	Institute of Astronomy, National Tsing Hua University, Hsinchu 30013, Taiwan
           \and
	Sandro Mereghetti \at
	INAF, Istituto di Astrofisica Spaziale e Fisica Cosmica, Via A. Corti 12, I-20133 Milano, Italy
           \and
	Roberto Mignani \at
	INAF, Istituto di Astrofisica Spaziale e Fisica Cosmica, Via A. Corti 12, I-20133 Milano, Italy; 
	Janusz Gil Institute of Astronomy, University of Zielona G\'ora, ul Szafrana 2, 65-265, Zielona G\'ora, Poland
           \and
	Antonios Nathanail \at
	Department of Physics, National and Kapodistrian University of Athens, Panepistimiopolis, GR 15783 Zografos, Greece
           \and
	Asaf Pe'er \at
	Department of Physics, Bar-Ilan University, Ramat-Gan 52900, Israel
           \and
	Silvia Piranomonte \at
	INAF, Osservatorio Astronomico di Roma, via Frascati 33, I-00078 Monte Porzio Catone (RM), Italy
           \and
	Piero Rosati \at
	Department of Physics and Earth Sciences, University of Ferrara, Ferrara, Italy
           \and
         Sandra Savaglio  \at          
       Physics  Department,  University  of  Calabria,  via  P.  Bucci, 87036 Rende, Italy
            \and           
	Fabian Sch\"ussler \at
	IRFU, CEA, Universit\'e Paris-Saclay, F-91191 Gif-sur-Yvette, France
            \and
	Olga Sergijenko \at
	Astronomical Observatory of Taras Shevchenko National University of Kyiv, Observatorna str. 3, Kyiv 04053, Ukraine; 
	Main Astronomical Observatory of the National Academy of Sciences of Ukraine, Zabolotnoho str. 27, Kyiv 03680, Ukraine
           \and
	Lijing Shao \at
	Kavli Institute for Astronomy and Astrophysics, Peking University, Beijing 100871, China; 
	National Astronomical Observatories, Chinese Academy of Sciences, Beijing 100012, China
           \and
	Sara Turriziani \at
	Physics Department, Gubkin Russian State University, 65 Leninsky Prospekt, Moscow 119991, Russian Federation
           \and
	Yuji Urata \at
	Institute of Astronomy, National Central University, Chung-Li 32054, Taiwan
           \and
	Maurice van Putten \at
	Physics and Astronomy, Sejong University, 98 Gunja-Dong Gwangin-gu, Seoul 143-747, Korea; 
	OzGrav-UWA, University of Western Australia, 35 Stirling Highway, M013, 6009 Crawley, WA, Australia
           \and
	Susanna Vergani \at
	GEPI, Observatoire de Paris, PSL University, CNRS, Place Jules Janssen, 92190 Meudon, France
	  \and
	Silvia Zane \at
	Mullards Space Science Laboratory, University College London, Holmbury St Mary, Dorking, Surrey, RH56NT, UK
           \and
	Bing Zhang \at
	Department of Physics and Astronomy, University of Nevada Las Vegas, Las Vegas, NV 89154, USA
}

\date{Received: date / Accepted: date}

\maketitle

\begin{abstract}
Multi-messenger astrophysics is becoming a major avenue to explore the Universe,
with the potential to span a vast range of redshifts. The growing synergies between different probes is opening new frontiers, which promise profound insights into several aspects of fundamental physics and cosmology. In this context, THESEUS will play a central role during the 2030s in detecting and localizing the electromagnetic counterparts of gravitational wave and neutrino sources that the unprecedented sensitivity of next generation detectors will discover at much higher rates than the present. 
Here, we review the most important target signals from multi-messenger sources that THESEUS will be able to detect and characterize, discussing detection rate expectations and scientific impact. 

\keywords{multi-messenger astrophysics \and gamma-ray burst \and compact binary merger \and kilonova \and X-ray sources \and neutrino sources}
\end{abstract}

\section{Introduction}

The breakthrough discoveries of the last few years have demonstrated the great scientific potential of gravitational wave (GW) astronomy and of multi-messenger astrophysics with GW and neutrino sources.
Since the first detection of GWs in 2015 from coalescing binary black hole (BH-BH) systems \cite{LVC-BBH1,LVC-BBH2}, tens of additional stellar-mass black hole coalescences \cite{LVC-GWTC-2} as well as two confirmed binary neutron star (NS-NS) mergers and at least one possible NS-black hole (NS-BH) merger \cite{LVC-BNS,LVC-BNS2,LVC-NSBH} have been detected so far with Advanced LIGO \cite{aLIGO} and Advanced Virgo \cite{AdVirgo}.
These observations likely represent only the tip of the iceberg and have confirmed the expectation that compact binary coalescences (CBCs) would represent the most common GW sources at the high frequencies where ground-based GW detectors are sensitive (i.e.~from $\sim10$\,Hz up to a few kHz).
At such frequencies, there are also other potentially detectable \\ GW sources, including core-collapsing massive stars as well as rotating and/or bursting NSs, whose output in GWs is however more uncertain with respect to CBCs (e.g., \cite{Abbott2020SN,Abbott2019cont}). 

All these high-frequency GW sources (possibly including stellar-mass BH-BH coalescences in rare circumstances; e.g., \cite{Bartos2017}) are expected to emit a variety of bright electromagnetic (EM) signals over the entire spectrum, from radio to gamma-rays (see Sections \ref{mergers}, \ref{otherGW}), offering opportunities for a multi-messenger investigation. 
The first GW detection of a NS-NS coalescence on August 17th 2017 \cite{LVC-BNS}, accompanied by the observation of the short gamma-ray burst (GRB) 170817A \cite{LVC-GRB}, the optical/infrared kilonova AT2017gfo, and further X-ray, optical, infrared, and radio emission (\cite{LVC-MMA} and refs.~therein), provided a first striking example of what can be accomplished by combining together the information from these two distinct channels (see also Section \ref{mergers}).

During the next few years, the aLIGO and AdVirgo will reach their design sensitivity and together with the first underground GW interferometer KAGRA in Japan \cite{Somiya2012}, which recently joined the network, they will ensure an increase in CBC detection rates and an improvement in source localization \cite{LVK-LRR-2020}.
By the end of the 2020s, further upgrades on aLIGO (A+ and Voyager \cite{Reitze2019}) and AdVirgo (Virgo+) are planned to be completed and the GW sky will be routinely monitored with the final second-generation (2G) GW detector network, composed by five interferometers with the addition of LIGO-India, a clone of the two LIGO detectors \cite{LVK-LRR-2020}. 
The distances up to which CBCs will be detected by the 2G network will go from few hundreds of Mpc to few Gpc \cite{LVK-LRR-2020}. Within such distances, the expected 2G network detection rate of NS-NS coalescences, i.e.~the most promising multi-messenger sources, could be as high as $80$/yr \cite{LVK-LRR-2020}. Nonetheless, joint short GRB observations by current and future high-energy missions that will be operational during the 2020s as, e.g., Swift \cite{Gehrels2004-Swift}, Fermi \cite{Atwood2009-FermiLAT,Meegan2009-FermiGBM}, INTEGRAL \cite{Winkler2003-INTEGRAL}, or SVOM \cite{Wei2016-SVOM} are still expected to be rare and likely less than one per year for geometrical reasons\footnote{Only a small fraction of NS-NS coalescences will be face-on, i.e.~with their orbital angular momentum nearly directed along the line of sight (within a few degrees). Even assuming a very high jet production efficiency from such systems, most of the corresponding short GRBs will be beamed away from us. The possible detection of ``off-axis'' or ``misaligned'' short GRBs, like in the case of GRB\,170817A, will remain limited to very near (and very rare) events.}. 
\begin{figure}[!t]
  \centering
  \includegraphics[width=1.0\linewidth]{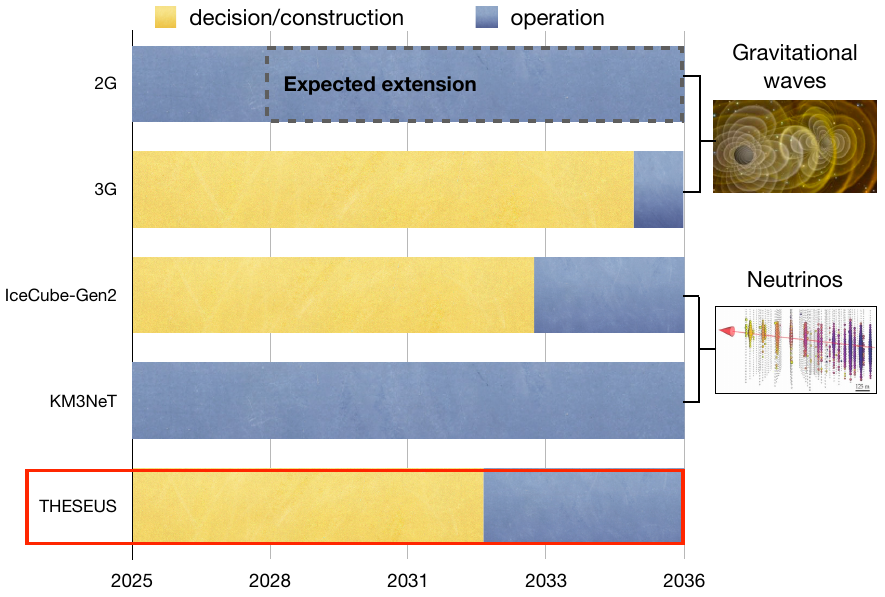}
  \caption{THESEUS nominal 4-year operation window along with the current timeline of major facilities for neutrino and GW observations by the end of the 2020s and the first half of 2030s, namely: the second-generation (2G) GW interferometer network (\cite{LVK-LRR-2020}; GWIC Roadmap \cite{GWIC-roadmap}, \url{https://gwic.ligo.org/}), the third-generation GW interferometers Einstein Telescope (ET; \cite{Punturo2010-ET}, \url{et-gw.eu}) and Cosmic Explorer (CE; \cite{Reitze2019}), and the neutrino observatories IceCube-Gen2 \cite{Aartsen2014} and  KM3NeT (ESFRI Roadmap 2018, \url{http://roadmap2018.esfri.eu}).}
  \label{fig1}
\end{figure}

About ten times more sensitive, third generation (3G) ground-based GW interferometers, such as the Einstein Telescope (ET; e.g., \cite{Punturo2010-ET,Maggiore2020}) and Cosmic Explorer (CE; e.g., \cite{Reitze2019}), are being planned for operation starting in the first half of the 2030s, allowing us to observe CBCs at distances nearly ten times farther with respect to the 2G network (see Figure~\ref{fig1}). 
This will significantly boost the detection rates for CBCs and, 
at the same time, greatly enhance the detection chances for the other types of fainter GW sources in the nearby Universe.
However, the next generation of ground-based GW interferometers will have relatively poor sky localisation capabilities for the vast majority of detected GW sources, implying serious difficulties in the identification of the EM counterparts. 
For instance, a network composed by ET and all the 2G detectors will localize within a sky area below 100\,deg$^2$ only $10-20$\% of NS-NS coalescences at $z\simeq0.3$ (e.g., \cite{Chan2018,Maggiore2020}).

Key discoveries have also been made in neutrino astronomy during the last decade, with at least two major results: (i) a diffuse flux of astrophysical very-high-energy neutrinos (10 TeV-10 PeV) detected by IceCube \cite{Aarsten2013}, the origin of which is still to date unknown (e.g., \cite{Aartsen2017}); (ii) the possible identification of a neutrino cosmic source with the blazar TXS0506+056 \cite{Aartsen2018}, which adds  to the only two previously known sources of neutrinos, both belonging to the Local Group environment, i.e.~the Sun and the supernova SN\,1987A. 
Among the most promising candidates for the diffuse neutrinos, GRBs, AGNs, and star bursting galaxies are of particular relevance and, for those, multi-messenger observations will be crucial to achieve the sensitivity level required by detection, thanks to the possibility of exploring spatial correlations as well as temporal coincidences in the case of transient events (see Section~\ref{neutrino}). 
Looking ahead towards the future multi-messenger campaigns, larger volume detectors are being planned, in particular gigaton water Cherenkov telescopes such as KM3NeT in the Mediterranean Sea \cite{Adrian2016} and IceCube-Gen2 at the South Pole \cite{Aartsen2014} (see also \cite{Agostini2020,Avrorin2018}).
In the early 2030s, these detectors will be completed, accessing the level of fluxes expected from cosmic sources (Figure~\ref{fig1}). 
Their sky localisation capabilities will however remain rather limited (e.g., \cite{Santander2018} and refs.~therein). 

In order to maximise the science return of the multi-messenger investigations during the 2030s, it will be essential to have a facility that can both (i) detect, localize, and disseminate the EM counterpart signals independently from the GW/neutrino events and, at the same time, (ii) rapidly cover with good sensitivity the large compatible sky areas provided by GW or neutrino detections. Moreover, given the lack of precise knowledge about the properties of various EM counterparts of both GW and neutrino sources, (iii) a large spectral coverage is another essential capability. 
These combined requirements are uniquely fulfilled by the Transient High-Energy Sky and Early Universe Surveyor (THESEUS).\footnote{We refer to the THESEUS Assessment Study Report (\url{https://sci.esa.int/s/8Zb0RB8}) for a general introduction on the space mission, the on-board instruments (XGIS, SXI, IRT), and the key scientific objectives.} 

THESEUS will allow us to monitor the transient sky with a number of advantages with respect to previous missions, yielding a significant step forward in our ability to investigate the multi-messenger Universe:
\begin{itemize}
\item A large fraction of poorly localised multi- messenger sources will be independently discovered with the THESEUS XGIS and SXI within one orbit, due to the unprecedented combination of large field-of-view (XGIS: 2\,sr in the $2-150$\,keV energy range and $>\!4$\,sr at $>\!150$\,keV; SXI: 0.5\,sr) and grasp (i.e.~the product of effective area and FoV) of these instruments. This will enable independent triggers on EM counterparts of numerous GW/neutrino sources, as it was the case for GRB\,170817A triggered by Fermi/GBM independently from the GW detection of the same source. At the same time, XGIS and SXI will provide fairly accurate localisation ($<\!15'$), which is a missing feature in Fermi/GBM. This will allow for follow-up observations with the onboard 0.7\,mt IR telescope (IRT) as well as other space and ground-based narrow field instruments. Sky coordinates can be disseminated to the astronomical community within minutes.
\item In case of detection of the NIR/optical counterpart by IRT, in response to an SXI/XGIS trigger, disseminated sky coordinates will be accurate at the arcsecond level. This fundamental input will make it possible to trigger deeper follow-up observations with the very large ground- and space-based telescopes available in the early 2030s, such as SKA, CTA, ELT, or Athena, which will further boost the scientific return in terms of GW and/or neutrino source characterization.
\item The high cadence spectral observations across the wide range 0.3\,keV$-$20\,MeV (SXI + XGIS), possibly with additional NIR coverage (IRT), will represent a great advantage for the identification and characterization of the diverse EM counterparts of GW and neutrino sources with respect to other all sky monitors that are limited to a narrower band, such as the forthcoming Chinese mission, Einstein Probe (0.3$-$4\,keV).
\item In response to THESEUS triggers, the search for sub-threshold events in GW and neutrino archival data will also be enabled (e.g., in case of a GRB trigger). Such a strategy has been already pursued by the current LIGO-Virgo Collaboration for a number of detected GRBs (e.g., \cite{Abbott2017-GRBsearch}).
\end{itemize}

The next Sections describe the main expected EM counterparts that THESEUS will be able to detect in synergy with the future GW and neutrino facilities, both in Survey mode and via Target of Opportunity programs. 
In Section~\ref{mergers}, we focus on the EM counterparts of NS-NS and NS-BH mergers, representing the most promising GW and multi-messenger sources. Section~\ref{otherGW} is devoted to GW sources with detectable EM counterparts other than merging compact binaries. Then, we complete the discussion on EM counterparts that THESEUS will be able to detect independently from an external trigger (Survey mode) with the most promising multi-messenger neutrino sources (Section~\ref{neutrino}). 
EM counterparts detectable by THESEUS in response to external triggers are discussed in Section~\ref{external}, while we draw our conclusions in Section~\ref{concl}.

\section{Electromagnetic counterparts from NS-NS and NS-BH mergers}
\label{mergers}

NS-NS and NS-BH mergers are among the most promising high-frequency GW sources (for ground-based interferometers) from which we expect a variety of detectable EM counterparts. 
From short GRBs to other X-ray and IR signals accompanying these merger events, we discuss here the main EM counterparts that THESEUS will be able to detect.

\subsection{Short gamma-ray bursts}
\label{SGRB}

\begin{figure}[!t]
  \centering
  \includegraphics[width=1.0\linewidth]{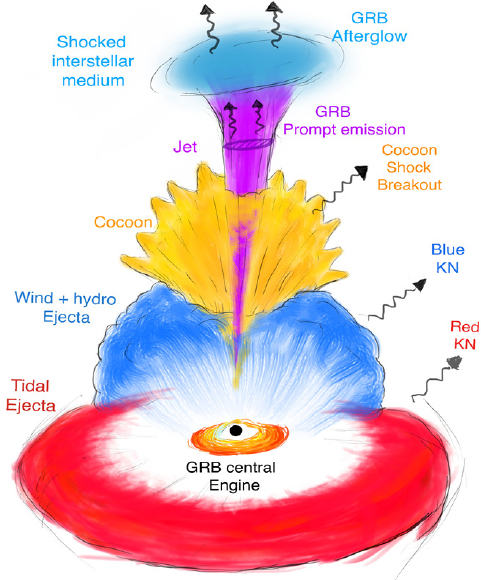}
  \caption{Schematic cartoon depicting the different emitting regions responsible for the EM counterparts of the multi-messenger event GW170817/GRB\,170817A, based on our current understanding of the physical processes accompanying the 2017 NS-NS merger. [From \cite{Ascenzi2021}]}
  \label{YB_2-12}
\end{figure}

The NS-NS merger detected with LIGO and Virgo on August 17, 2017 (GW170817) and its associated short \\ GRB\,170817A was the first direct evidence of the progenitor of a short GRB as a compact binary merger system 
\cite{LVC-GRB,Goldstein2017,Savchenko2017,Troja2017,Margutti2017,Hallinan2017,Alexander2017,Mooley2018a,Lazzati2018,Lyman2018,Alexander2018,Mooley2018b,Ghirlanda2019} (see, e.g., \cite{Nakar2020} for a review), which confirmed several indirect pieces of evidence collected in the last decade (e.g., \cite{Tanvir2013,Berger2014}). The afterglow properties of \\ GRB\,170817A also confirmed the formation of a relativistic, narrow jet (half-opening angle of about $2-4$\,deg \cite{Mooley2018b,Ghirlanda2019}) after the NS-NS merger, a result that theoretical studies and MHD simulations could not fully predict. It was also the first short GRB viewed from outside the core of the jet (i.e.~the cone with very high Lorentz factor), as demonstrated by the rising and then slowly decaying afterglow. The viewing angle was estimated to be around $15-30$\,deg away from the direction of propagation of the highly relativistic jet core \cite{Mooley2018b,Ghirlanda2019}. Such a lateral view, allowed to identify the observed gamma-ray emission as directly originating from the mildly relativistic cocoon that formed around the jet core via the interaction of the incipient jet itself with the surrounding material ejected during and after the NS-NS merger. Figure~\ref{YB_2-12} depicts our current understanding of NS-NS merger emitting regions, as gathered from the single multi-messenger observation of the August 2017 event.

The above results clearly show how the detection of short GRBs is of crucial relevance for multi-messenger astrophysics and underline the fundamental role of THESEUS in ensuring short GRB observations during the 2030s, when the current and future space missions as Fermi, Swift, or SVOM are not guaranteed and, at the same time, both 2G and 3G GW interferometers are expected to be operational. 

\begin{figure*}[!t]
  \centering
  \includegraphics[width=0.999\linewidth]{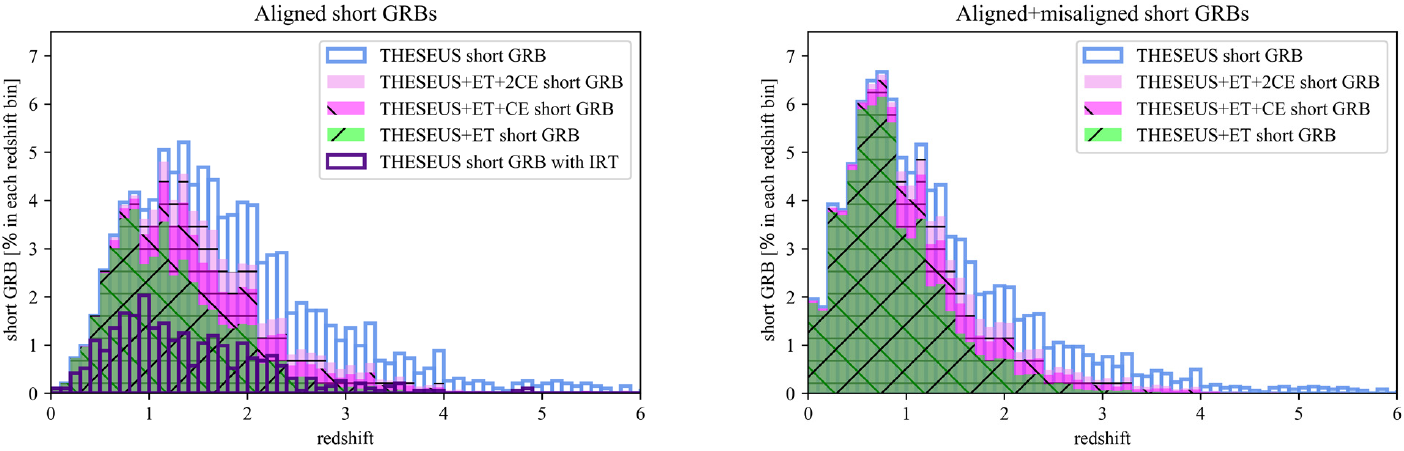}
  \caption{{\it Left:} The redshift distribution of well-localized aligned short GRBs (blue) from XGIS and SXI and those detected also with IRT (25\%, indigo). Joint short GRB+GW detections are obtained by considering, at each redshift, the GW detection efficiency for NS-NS mergers by ET (green, 46\% of THESEUS short GRBs), the ET+CE network and ET+2\,CEs network (magenta and pink, 62\% and 73\%), respectively. {\it Right:} Same as the left panel where misaligned short GRBs are also included (see text). Including misaligned events not only increases the total number of THESEUS short GRBs, but also boosts the fraction of events with a joint EM+GW detection, leading to 63\% for ET, 76\% for ET+CE and 83\% for ET+2\,CEs. }
  \label{YB_2-14}
\end{figure*}

During its nominal mission lifetime, THESEUS/XGIS and SXI are expected to detect and accurately ($<\!15'$) localize $\simeq\!40$ short GRBs ($\simeq\!12/$yr assuming 3.45 years of scientific observations) inside their imaging field of view, plus numerous short GRBs at higher energies ($>\!150$\,keV) with coarse or no sky localization. 
These numbers are obtained from simulations of a realistic observational sequence of THESEUS, considering all observational constraints, in response to a random set of short GRB triggers based on the population model of \cite{Ghirlanda2016}.\footnote{For more details, see the THESEUS Assessment Study Report (\url{https://sci.esa.int/s/8Zb0RB8}).} Such a population model is built on short GRBs observed before GRB\,170817A with Swift and Fermi, for which the line of sight falls inside the narrow core of the corresponding jet (i.e.~``aligned'').\footnote{We note that other population models for aligned short GRBs exist in the literature (e.g., \cite{Wanderman2015}).}
Figure \ref{YB_2-14} (left panel) shows the redshift distribution of these short GRBs (blue line). Joint short GRB+GW detections are also shown, obtained by considering, at each redshift, the GW detection efficiency for NS-NS mergers. In these computations, three scenarios for the 3G GW interferometers have been considered: 1) ET alone, 2) ET plus one CE (in USA), 3) ET plus two CEs (one in USA and one in Australia). The expected numbers of short GRBs detected and localized with THESEUS and detected also by 2G and 3G interferometers are summarized in Table~\ref{TAB_2-3}. These conservative numbers are robust and based on the Mission Observation Simulator (MOS) results$^3$ and state-of-the-art NS-NS merger simulations for the GW detection efficiency estimates.
\begin{table*}[t!]
\begin{tabular*}{\textwidth}{@{\extracolsep{\fill}} ccccc}
\hline
GW detectors & THESEUS+GW detectors & aligned short GRB+GW & aligned \& misaligned \\
 & plausible joint observation time & detections & short GRB+GW detections \\
\hline
\hline
2G network & 3.45 yr & $\sim\!0.04$ & 1.8 \\
\hline
ET & 1 yr (3.45 yr) & 5.6 (19.2) & 13 (46) \\
\hline
ET+CE & 1 yr (3.45 yr) & 7.4 (25.7) & 16 (55) \\
\hline
ET+2\,CEs & 1 yr (3.45 yr) & 8.7 (30.1) & 18 (61) \\
\hline
\end{tabular*}
\caption{\label{TAB_2-3} 
Expected number of joint prompt GW+EM detections of NS-NS mergers/short GRBs for THESEUS and different GW detectors, assuming 1 or 3.45 years of joint observations. Number estimates of aligned short GRB+GW detections take into account the redshift distribution of THESEUS short GRBs from the MOS, and the NS-NS merger detection efficiency at each distance/redshift as predicted for the different GW detectors, assuming SNR$=$8. Number estimates of aligned plus misaligned short GRBs+GW detections take into account also the maximum viewing angle for misaligned short GRB detection at each distance/redshift (see text and Figs.~\ref{YB_2-14}, \ref{YB_2-13}).}
\end{table*}

\begin{figure}[!t]
  \centering
  \includegraphics[width=1.0\linewidth]{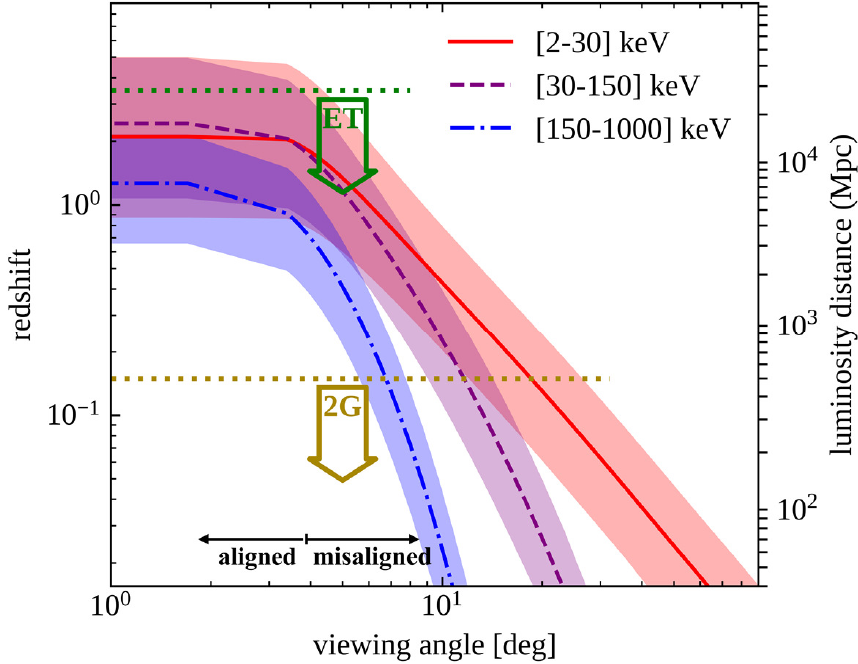}
  \caption{Maximum distance/redshift for detecting with THESEUS the prompt emission of a short GRB like 170817A versus the viewing angle, depending on the energy band (red, violet, and blue lines/stripes).
Calculations are based on \cite{Salafia2019} and employ a series of simplifying assumptions.}
  \label{YB_2-13}
\end{figure}

By considering the possibility to observe short GRBs also outside the solid angle of the narrow jet core, the number of potential detections can sensibly increase. Indeed, the misaligned view of GRB\,170817A enabled us for the first time to quantify how the high-energy prompt emission becomes gradually softer and less energetic as the viewing angle increases (with respect to the jet axis). 
As a result, it has been possible to estimate, for events similar to GRB\,170817A, the maximum viewing angle at which a given instrument could detect the prompt emission depending on distance (see Figure~\ref{YB_2-13}).\footnote{We refer here to the GRB\,170817A jet angular structure as inferred in \cite{Ghirlanda2019}. We note that there are also different angular structures compatible with the observations (e.g., \cite{Ioka2019}).} 
Based on such an estimate, the unique capabilities of THESEUS offer excellent prospects for detecting the prompt emission from misaligned short GRBs within the relatively small distance reach of GW detectors. In particular, for NS-NS mergers detected by 2G interferometer network, the GRB\,170817A-like prompt emission would be observable up to $10-30$\,deg, depending on the energy band (Figure~\ref{YB_2-13}), corresponding to a detection rate increased by almost a factor of 50 with respect to the result for aligned events only (Table~\ref{TAB_2-3}). At the typical distance reached by a 3G detector such as ET, the prompt emission would still be observable by THESEUS up to order $\sim\!10$\,deg, more than doubling the joint detection rate (Table~\ref{TAB_2-3}).

Building statistically relevant samples of short GRBs for which coincident GW observations will be available (Table~\ref{TAB_2-3}), which is a unique capability of THESEUS, will allow for unprecedented investigations on the nature of compact binary mergers. 
Fundamental open questions on the nature of CBC sources and short GRB central engines that THESEUS will allow us to solve in synergy with the next generation GW interferometers include the following: 
\begin{itemize}
\item {\bf How frequent is relativistic jet formation in NS-NS and NS-BH mergers?} THESEUS will allow for the detection of at least a few to about 10 or more short GRBs associated with GW-detected NS-NS/NS-BH mergers. The association of a short GRB with NS-NS/NS-BH mergers unambiguously brought us the proof of the formation of a relativistic jet. Along with detections, THESEUS will also allow for confident non-detections in case of face-on mergers without a short GRB (based on the binary system inclination extracted via the GW signal). 
\item {\bf What is the jet launching mechanism in NS-NS/NS-BH mergers?} The time delay between the GW merger epoch and the GRB peak flux is a powerful diagnostic indicator for the jet launching mechanism (e.g., \cite{LVC-GRB,Zhang2019,Gill2019,Lazzati2020}), which is still a matter of debate (e.g., \cite{Rezzolla2011,Paschalidis2015,Just2016,Ruiz2016,Ciolfi2020a}). The significant number of short GRBs observed by THESEUS in synergy with GW detectors will allow us to uniquely characterize this important parameter and highlight differences between NS-NS and NS-BH systems. 
\item {\bf What is the nature of the short GRB central engine and the origin of the still unexplained extra-features (e.g., ``Extended Emission'', ``Plateaus'')?} \\ For short GRBs detected by THESEUS, the subsequent X-ray emission will be observed via the on-board SXI and/or by communicating the accurate sky localization to X-ray telescopes such as Athena. In presence of a coincident GW detection, a combined analysis will be possible, shedding light on the nature of the merger remnant (i.e.~accreting BH or massive NS; e.g., \cite{Ciolfi2015,Rezzolla2015,Ciolfi2018,Ai2020,Ciolfi2020a}). This unprecedented collection of information will also unveil the origin and statistical properties of puzzling X-ray features like the Extended Emission and the X-ray plateaus (see Sections \ref{EE}, \ref{plateau}). 
\item {\bf Do jets have a universal structure and are there any systematic differences between NS-NS and NS-BH \\ mergers?} The afterglow properties of short GRBs viewed from outside the core of the jet strongly depend on the jet structure (and in particular the energy and Lorentz factor angular distribution around the jet axis). THESEUS will detect and localize down to arcmin level several misaligned short GRBs (Table~\ref{TAB_2-3}). The afterglow profile of the brightest nearby sources will be monitored with SXI and IRT (see Section \ref{afterglow}). Moreover, synergy with powerful facilities, such as the contemporaneous mission Athena, will allow for deep and long afterglow monitoring. 
\item {\bf Do jets have a universal structure and are there any systematic differences between NS-NS and NS-BH \\ mergers?} The afterglow properties of short GRBs viewed from outside the core of the jet strongly depend on the jet structure (and in particular the energy and Lorentz factor angular distribution around the jet axis). THESEUS will detect and localize down to arcmin level several misaligned short GRBs (Table~\ref{TAB_2-3}). The afterglow profile of the brightest nearby sources will be monitored with SXI and IRT (see Section \ref{afterglow}). Moreover, synergy with powerful facilities, such as the contemporaneous mission Athena, will allow for deep and long afterglow monitoring. 
\item {\bf What is the role of merging NS-NS and NS-BH systems in the chemical enrichment of the Universe?} Kilonova observations provide crucial information on the r-process element formation accompanying these events, which is a fundamental open problem. Moreover, the overall contribution to the r-process element abundances relative to the one from supernovae (SNe) remains unclear.  
THESEUS accurate sky localization of several NS-NS/NS-BH mergers will allow for kilonova detection and characterization through the follow-up with the onboard NIR telescope and/or through ground-based follow-up campaigns (see Section \ref{kilonova}). 
\end{itemize}

Besides the short GRB prompt emission, other EM signals directly related to short GRBs are expected to be detected with THESEUS jointly with GW observations of CBC events. These additional EM counterparts are described in the following Sections and include the well-known jet afterglows as well as the so-called ``Extended Emission'' and ``X-ray Plateaus'' often observed in short GRB events, whose origin is still matter of debate.  
Extended Emission and X-ray Plateaus, never detected without the prompt short GRB emission, are of particular interest as they (i) might be significantly less collimated with respect to the latter and as such observable from a larger fraction of GW events, and (ii) could provide fundamental clues on the nature of the post-merger remnant.
\begin{figure*}[!t]
  \centering
  \includegraphics[width=0.99\linewidth]{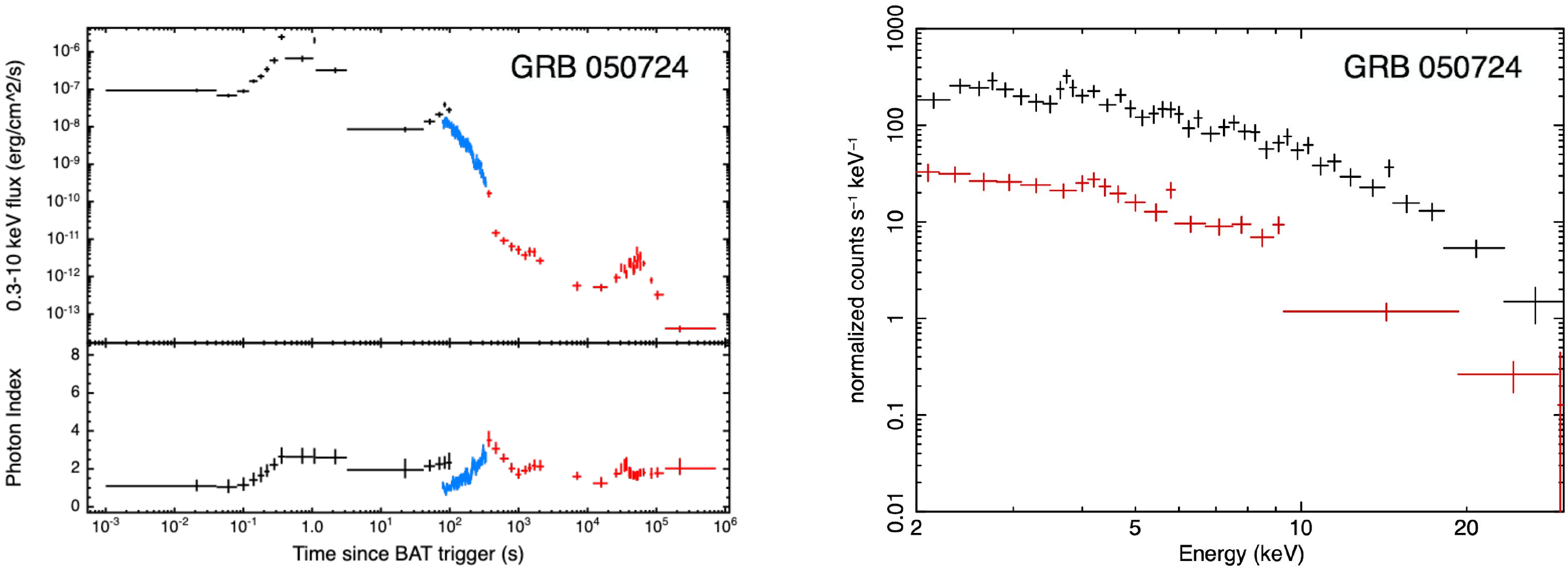}
  \caption{{\it Left:} Prototype of short GRB with Extended Emission (EE), GRB\,050724 at z$=$0.257, detected with Swift/BAT \cite{Barthelmy2005a}. [Figure produced via the Swift Burst Analyser \cite{Evans2010}] ~~{\it Right:} Simulations of the 2$-$30\,keV spectrum of GRB\,050724 obtained with XGIS (assuming 15\,deg off-axis detector calibration), where the main hard spike (black) and EE (red) are detected at 35 sigma with 3\,s of exposure and 22 sigma with 100\,s of exposure, respectively.}
  \label{EEobs}
\end{figure*}

\subsubsection{Extended Emission}
\label{EE}

A fraction of short GRBs, immediately after the hard spike, shows a softer and prolonged emission (``Extended Emission'', hereafter EE) lasting a few tens up to hundreds of seconds \cite{Norris2006}. Past attempts to quantify the fraction of short GRB with EE led to a wide range of values that goes from 2\% up to 25\%, depending also on the sensitivity band of the gamma-ray detector used for the classification \cite{Norris2010,Bostanci2013,Kaneko2015}. A recent systematic analysis of Swift XRT and BAT data of a sample of 65 short GRBs (6 times larger than past studies, \cite{Kisaka2017} suggests the presence of a severe bias against the lack of an X-ray view of the prompt emission, with a true fraction of short GRBs accompanied by EE of more than 75\%.

A prototype of short GRBs with EE is GRB\,050724 at $z=0.26$ (Figure~\ref{EEobs}). Simulations of this burst show that THESEUS could have clearly detected both the main hard spike and the EE component with XGIS as well as characterized its spectrum. Further simulations over a sample of 8 short GRBs with EE at known redshift \cite{Kaneko2015} show that EE can be detected up to $z\!\simeq\!2$ and in some cases the detection significance of the EE component with XGIS is even higher than the detection significance of the main hard spike.

\begin{figure}[!h]
  \centering
  \includegraphics[width=1.0\linewidth]{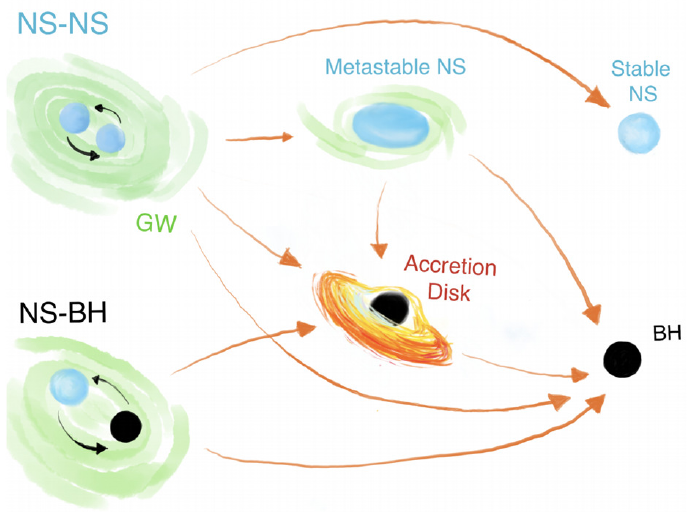}
  \caption{Schematic cartoon depicting the possible remnant formation channels of NS-NS and NS-BH mergers. [From \cite{Ascenzi2021}].}
   \label{YB_2-16}
\end{figure}

\begin{figure}[!t]
  \centering
  \includegraphics[width=1.0\linewidth]{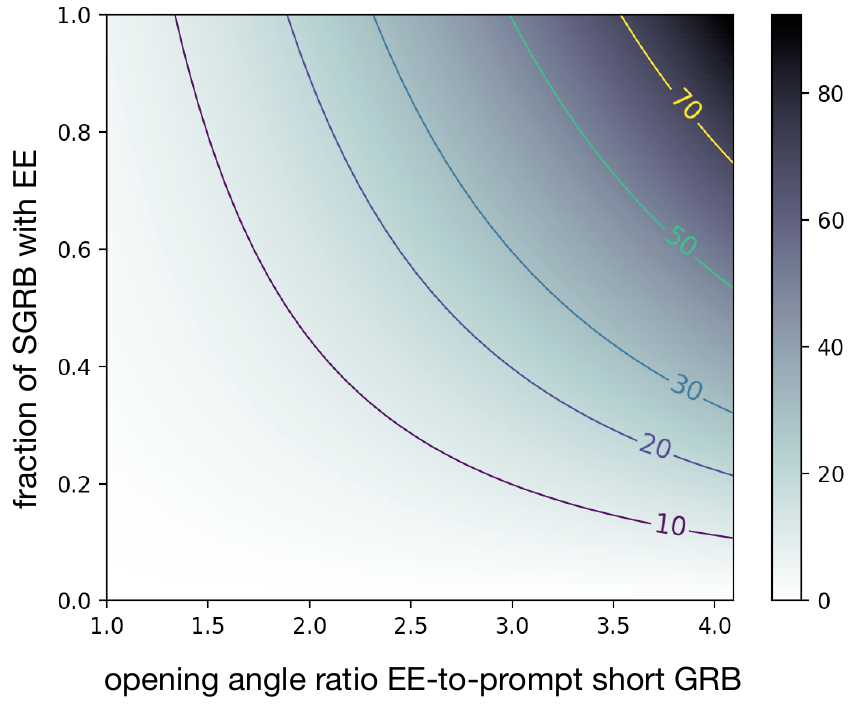}
  \caption{Number of EE signals that THESEUS can detect in 1 year in combination with ET GW detections, depending on the fraction of short GRBs accompanied by EE and on the ratio between the characteristic opening angle of the EE and that of the short GRB jet core.}
   \label{N_EE}
\end{figure}

The physical interpretation of the EE is still unclear. The long EE duration (order $10-100$\,s) challenges the leading scenario envisaging an accreting BH as the short GRB central engine and supports the formation of a long-lived spinning down massive NS remnant (e.g., \cite{Bucciantini2012}; see also Figure~\ref{YB_2-16}). In this alternative scenario, the EE is expected to be much less collimated with respect to the main spike.
For this reason, the EE can also represent a possible ``short GRB-less'' EM counterpart of NS-NS mergers, which can in turn further boost the number of EM counterparts of GW sources that THESEUS will be able to catch. 
As an illustrative example, Figure \ref{N_EE} shows the expected number of EE signals that THESEUS can detect in 1 year in combination with ET GW detections. This number depends on the two still uncertain parameters, namely the fraction of short GRBs with EE and the characteristic opening angle of the EE. For instance, by assuming a fraction of short GRBs with EE of 50\% and an EE half-opening angle four times larger than the main hard spike (assuming a jet half-opening angle of a few degrees, this would correspond to $10\!-\!15$\,deg), THESEUS would detect about 45 EE signals with a GW counterpart observed by ET. The largest fraction of these events will be ``short GRB-less'', thus adding to the overall number of multi-messenger detections enabled by THESEUS.

\subsubsection{X-ray plateaus}
\label{plateau}

The soft X-ray afterglow lightcurve of GRBs is often characterized by an initial steep decay, followed by a rather shallow decay phase (so-called ``plateau'' phase) which can extend up to several thousands of seconds. According to over 15 years of observations by Swift, a large fraction of all short GRBs ($\approx\!50$\%) may be accompanied by an X-ray plateau. 
A common interpretation for the X-ray plateaus is based on an external shock emission sustained by energy injection from an active central engine that can either be an accreting BH or a highly magnetized NS (see also Section~\ref{SDPT}). In the latter case, the plateau emission can be poorly collimated or even nearly isotropic (e.g., \cite{Siegel2016a,Siegel2016b}). According to an alternative interpretation, both the steep decay and the plateau are instead due to high-latitude emission (HLE) produced from a structured jet whose energy and bulk Lorentz factor gradually decrease with the angular distance from the jet symmetry axis (e.g., \cite{Oganesyan2020,Ascenzi2020}). This model predicts an X-ray emission that becomes fainter at larger polar angles and thus detectable, for a given distance, up to a maximum angle (see Figure~\ref{HLE}). 

\begin{figure}[!t]
  \centering
  \includegraphics[width=1.0\linewidth]{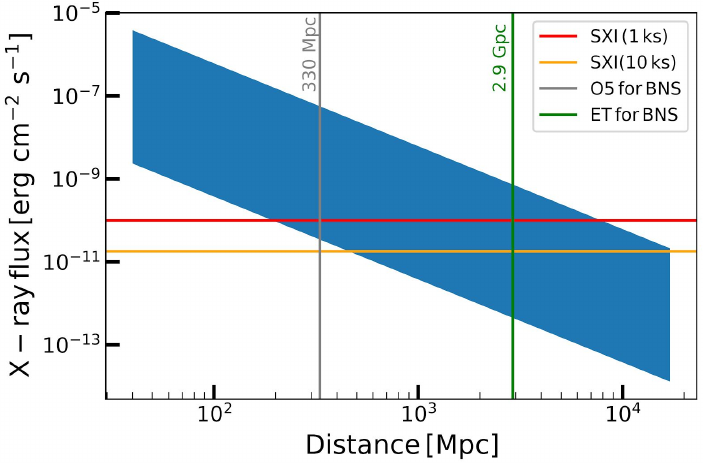}
  \caption{Expected flux range of X-ray plateaus associated with (aligned) short GRBs as computed from a selected sample of Swift/XRT events with known redshift and rescaled at different distances. Vertical lines indicate the typical distance at which a randomly oriented NS-NS merger can be detected with 2G network (330\,Mpc) or ET (2.9\,Gpc). }
   \label{plateaus}
\end{figure}

\begin{figure}[!t]
  \centering
  \includegraphics[width=1.0\linewidth]{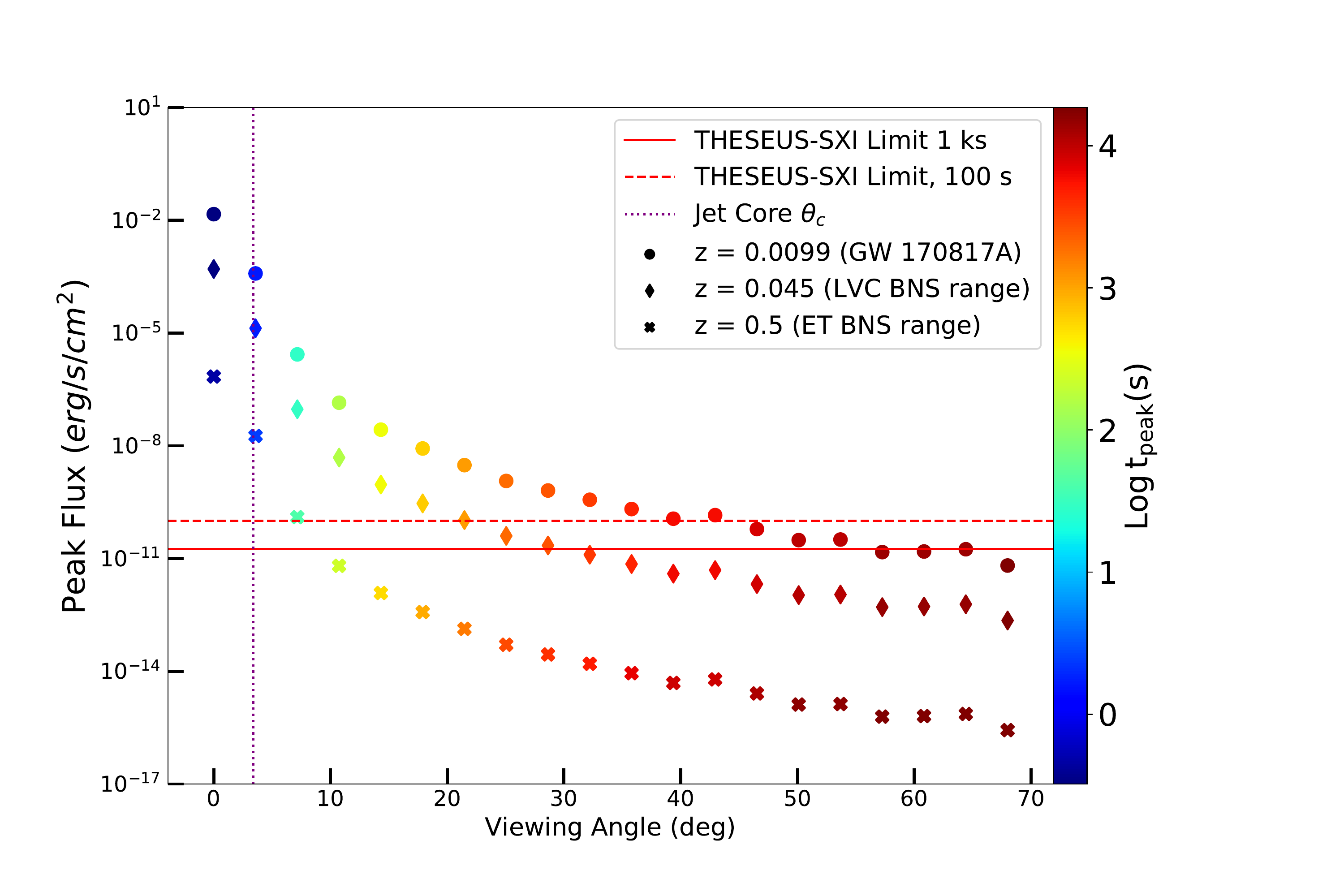}
  \caption{Predicted HLE peak fluxes at different viewing angles compared with THESEUS/SXI sensitivity with 100\,s and 1\,ks exposures (horizontal red dashed and solid lines, respectively), for a GRB\,170817A-like event placed at three different distances (circle, diamond, and cross markers). Color-coded is the peak time. Calculations are based on \cite{Oganesyan2020,Ascenzi2020} (see \cite{Ascenzi2020} for a similar figure referred to a different event).}
   \label{HLE}
\end{figure}

Figure~\ref{plateaus} shows the X-ray flux range spanned by the best observed plateaus associated with (aligned) short GRBs with redshift measurements, where the latter allows to rescale the flux itself with distance. A comparison with the sensitivity of the THESEUS SXI shows that such an instrument is perfectly suitable to catch this emission and, assuming an exposure of 1\,ks, would allow us to detect about 90\% (30\%) of all X-ray plateaus up to 330\,Mpc (2.9\,Gpc), which is the typical 2G (ET) GW detection distance for a randomly oriented NS-NS merger. 
For GRB\,170817A, the X-ray plateau lightcurve predicted by the HLE modelling is consistent with the non-detection by, e.g., Swift and MAXI. At the same time, the sensitivity of THESEUS/SXI, combined with the ability of THESEUS to trigger the burst and rapidly localize it, would have allowed for an early and confident detection.

So far, it has not been possible to disentangle the different interpretations of X-ray plateaus outlined above (and others; e.g., \cite{Lu2015,Beniamini2020}, as the predicted flux evolution can in any case fairly well reproduce the events  observed with \\ Swift/XRT (e.g., \cite{DallOsso2011,Rowlinson2013,Stratta2018,Oganesyan2020}). 
THESEUS will give us the opportunity to collect a statistically significant sample of plateau detections in synergy with GW observations and thus to constrain the emission model 
(possibly aided by the identification of the remnant nature, i.e.~BH vs.~NS, via the post-merger GW signal). In this respect, also the number of ``orphan'' X-ray plateaus detected (i.e.~without a prompt short GRB detection) will be revealing.

\subsubsection{Jet afterglows with SXI and IRT}
\label{afterglow}

The propagation of a GRB jet in the interstellar medium is known to produce a multi-wavelength afterglow signal, from X-rays to radio, via synchrotron emission at the forward shock \cite{Meszaros1997,Sari1998}. GRB\,170817A was the first short GRB viewed with line of sight significantly misaligned with respect to the jet axis and the properties of the observed afterglow radiation offered a unique chance to probe the angular jet profile. Taking this event as a reference, we can estimate the maximum distance at which the afterglow signal is above the detection sensitivity for the IRT and SXI instruments on-board THESEUS, depending on the viewing angle with respect to the jet propagation axis. 
In Figure~\ref{afterglow}, we show the result based on the power-law angular jet structure that best fits the observations according to \cite{Ghirlanda2019,Lamb2019} (see also \cite{Lamb2017,Lamb2018,Salafia2019}). A GRB\,170817A-like afterglow signal at 40 Mpc could be detected with SXI up to an inclination of $\simeq\!10$\,deg, with a peak emission time between a few hours and 1 day, and with IRT up to $\simeq\!20$\,deg, with peak emission time around 2 days. Going to a larger distance of 500 Mpc (nearly the maximum distance reach of 2G GW detectors for NS-NS mergers), SXI and IRT could detect a GRB\,170817A-like afterglow signal respectively up to $\simeq\!4.5$\,deg and $\simeq\!10$\,deg.

\subsection{Other CBC counterparts of interest for THESEUS}
\label{otherCBC}

\subsubsection{Spin Down Powered Transients}
\label{SDPT}

A potentially powerful nearly-isotropic emission is expected if a NS-NS merger produces a long-lived highly magnetized NS that does not collapse to a BH for as long as minutes, hours, or more (see Figure~\ref{YB_2-16}). Soft X-ray to optical transients can be powered by the NS EM spin-down emission reprocessed by the baryon-polluted environment surrounding the merger site (e.g., \cite{Zhang2013,Yu2013,Metzger2014,Siegel2016a,Siegel2016b}), which consists of a dense cloud of material expelled in the early post-merger phase (e.g., \cite{Ciolfi2019}). In soft X-rays, such spin-down powered transients (SDPTs) can last for a timescale of minutes to days and their expected luminosities are up to $10^{46}-10^{48}$\,erg/s \cite{Siegel2016a,Siegel2016b}, which would be reachable by SXI up to $0.9-9$\,Gpc with 1ks exposures.

One or more unambiguous detections of this type of emission after a NS-NS merger would indicate that the remnant is long-lived, allowing us to achieve significant constraints on the NS equation of state and other key properties of the remnant itself (e.g., \cite{Piro2017}). Moreover, it would provide crucial information to estimate the currently unknown fraction of mergers forming a long-lived NS remnant. Finally, it would clarify the possible connection with the extended emission and/or the X-ray plateaus of short GRBs. In the case of GW170817/GRB\,170817A, no evidence for a SDPT was found in the soft X-ray band. However, the first deep pointed observations at $\sim$keV photon energies only started as late as $\sim\!15$\,h after merger with Swift/XRT \cite{Evans2017}, and the earlier constraints provided by MAXI 4.6\,h after merger with a flux limit of $8.6\times10^{-9}$\,erg/(cm$^2$\,s) \cite{Sugita2018} were not able to exclude a SDPT. The combination of the THESEUS/SXI sensitivity at keV energies and its field of view about 10$^4$ times larger than Swift/XRT will offer much better prospects for a detection within minutes/hours after a GW trigger.

\subsubsection{Kilonovae with IRT}
\label{kilonova}

Neutron-rich matter released from NS-NS/NS-BH mergers undergoes rapid neutron capture (r-process) nucleosynthesis, leading to the formation of very heavy elements such as gold and platinum. This scenario likely provides a significant (if not dominant, compared to SNe) contribution to the observed abundances of rare heavy elements in the Universe. Radioactive decay of the newly-formed and unstable nuclei powers a rapidly evolving, nearly isotropic thermal transient known as a kilonova, the observation of which not only witnesses cosmic heavy element production, but can also probe the physical conditions during and after the merger phase (e.g., \cite{Metzger2019LRR}).

The first robust observation of a kilonova, following a few candidates (e.g., \cite{Tanvir2013,Jin2015}, was the optical and infrared counterpart of GW170817 (e.g., \cite{Arcavi2017,Coulter2017,Pian2017,Smartt2017,Kasen2017}; see also \cite{Metzger2019LRR} and refs.~therein), named AT2017gfo, discovered about 11 hours after the GW/GRB trigger via galaxy targeting inside the GW plausible sky area (\cite{LVC-MMA} and refs.~therein).\footnote{After AT2017gfo, the re-analysis of different events led to the identification of other likely kilonovae (e.g., \cite{Lamb2019kn}).} 
During the next decade, we may observe other kilonovae associated with nearby NS-NS and perhaps NS-BH mergers as well as kilonovae without GW counterparts. 
Then, in the 2030s, the IRT onboard THESEUS will also contribute to the search and localization of kilonova signals, in particular if associated with a detectable aligned or misaligned short GRB. 
As shown in Figure~\ref{kilonova}, the IRT can detect the full SED (Spectral Energy
Distribution) of a kilonova like AT2017gfo up to 320\,Mpc (180\,Mpc) with 600\,s (60\,s) of exposure, within one day from the merger epoch. At later times, the kilonova will be fading away, but IRT will still be capable to detect the source in each filter thus allowing to build spectral energy distribution up to 180 Mpc within a few days after the trigger. For nearby sources ($<\!40$\,Mpc), near-IR spectra can also be obtained.
\begin{figure*}[!t]
  \centering
  \includegraphics[width=0.9\linewidth]{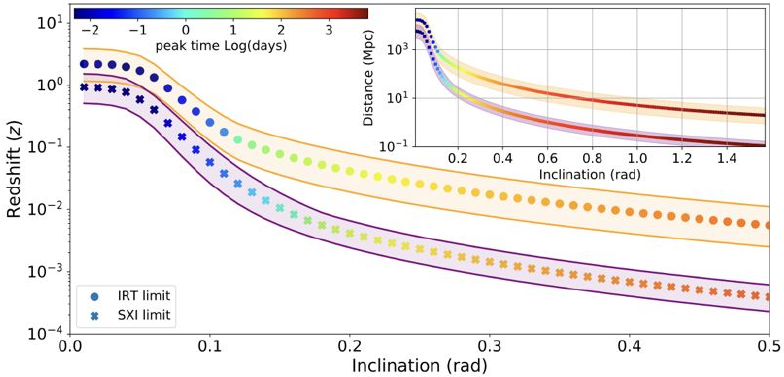}
  \caption{Maximum redshift/distance vs.~inclination for a THESEUS detection in the H-band (circles) and at 1 keV (X-rays) with IRT and SXI, respectively, for jet afterglow radiation assuming the GRB\,170817A power-law jet structure from \cite{Ghirlanda2019,Lamb2019}. Color-coded is the peak time. Calculations are based on \cite{Lamb2017,Lamb2018}.}
   \label{afterglow}
\end{figure*}

\subsection{CBC redshifts and prospects for $H_0$ measurement}
\label{H0}

In the last years, two main measurements of the Hubble constant $H_0$, obtained from Planck observations of the CMB and from the SNIa distance ladder, have come into significant tension with a steadily growing discrepancy, currently at more than 4 sigma level (e.g., \cite{Riess2021}). An independent, new measurement of $H_0$ would be of utmost importance in order to understand if the current discrepancy is due to possible systematics or is the sign of a cosmological crisis that requires new paradigms. The luminosity distance from the detection of GWs from CBCs and the measurement of their redshift through their EM counterpart has already proven to be a potential alternative probe for $H_0$ with the example case of GW170817 \cite{LVC-Hubble}. 

To solve the current tension, however, a precision level of the order of 1\% must be reached. 
In this context, THESEUS observations of a large number of short GRBs in synergy with 3G interferometers represent a unique opportunity. Simulations of NS-NS mergers observed with the 3G network along with an instrument like THESEUS/XGIS have been performed by \cite{Belgacem2019}. Their results predict a number of joint detections of $130-300$ in 10 years, from which $H_0$ could be measured with a precision of $0.2 - 0.4\%$ by assuming that a redshift can be measured for all events via either optical or X-ray spectroscopy: with this assumption, Figure~\ref{H0_plot} shows that, by rescaling these precision levels as $1/\sqrt{N}$, the goal of $\Delta H_0/H_0\sim1\%$ can be reached with $N\!\simeq\!15$ events jointly observed with the 3G network \\ (ET+2\,CEs) and $N\!\simeq\!25$ events jointly observed with ET only (the lower number of events providing $\Delta H_0/H_0\!\sim\!1\%$ with the ET+2\,CEs network with respect to ET only is due to the better parameter estimation with the former network). THESEUS can reach these detection numbers in 1-2 years of operations in synergy with 3G GW detectors (see Tab.~\ref{TAB_2-3}). 

However, as we learned from past observations, the redshift cannot be measured for all short GRBs due to host galaxy identification challenges. This will likely not affect the $\simeq25\%$ of THESEUS short GRBs detected with IRT since their sky localization to arcsec accuracy will enable unambiguous identification of the host galaxy and redshift measurement (in the vast majority of cases). For the remaining $\simeq75\%$ without IRT detection, the large number of galaxies contained in the XGIS or SXI error boxes for almost all short GRBs (i.e.~at distances $>\!50\!-\!100$\,Mpc) severely challenges the identification of the host galaxy if no transient optical afterglow is detected.
In order to quantify the chances to perform successful ground-based afterglow follow-up, we generated\footnote{Using the python module afterglowpy \cite{Ryan2020} that however does not take into account possible ``rebrightenings'' observed in several optical afterglows, the origin of which is not yet fully understood (e.g., \cite{Lamb2019kn}). Therefore, provided estimates are conservative.} 1000 optical aligned and misaligned afterglow synthetic light curves assuming GRBs with equivalent isotropic radiated energy $>\!10^{50}$\,erg and mean value $\simeq\!2\times10^{51}$\,erg and then compared with the magnitude limits of different telescopes that may operate in the era of THESEUS (in particular, we considered here LSST/VRO, the Liverpool Telescope, and GTC/OSIRIS).
Results show that, for short GRBs observed with viewing angle between 0 and 10 degrees with respect to the jet axis and with no IRT detection, $\simeq\!50\%$ will have a detectable optical afterglow (that unambiguously pinpoints the host galaxy) by providing a ground-based telescope follow-up reaction time of a few hours. With the same assumptions, $\simeq\!13\%$ of short GRB observed with viewing angle between 10 and 30 degrees with respect to the jet axis will have a detected optical afterglow. 
By taking into account these results, $H_0$ should be measured with $\sim 1$\% accuracy (at 1 sigma) with 1 yr of synergy with the ET+2\,CEs network and 3.45\,yr with ET (Figure~\ref{H0_plot}).
\begin{figure}[!t]
  \centering
  \includegraphics[width=1.0\linewidth]{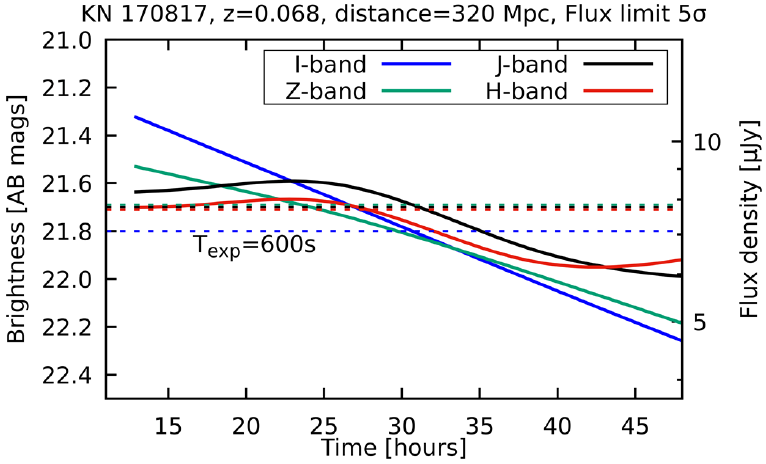}
  \caption{Multi-band light curves of the kilonova AT2017gfo (from \cite{Rossi2020}) compared with THESEUS/IRT sensitivity. THESEUS can detect a kilonova like AT2017gfo with 5 sigma up to 320\,Mpc in all bands with 600\,s exposure, within $1-2$\,days from the merger epoch. Near-IR spectra ($H\!<\!17.5$, 1800\,s) can be obtained for the most nearby sources (i.e.~$<\!40$\,Mpc).}
   \label{kilonova}
\end{figure}

\section{Other GW sources}
\label{otherGW}

\subsection{Core-collapse of massive stars}
\label{CCSNe}

Beside CBCs, core-collapse supernovae (CCSNe) represent another type of GW sources that are of great interest for the astrophysics community. However, contrary to the CBC case, their expected GW emission is highly uncertain as it strongly depends on the rather unknown SN explosion mechanism (e.g., \cite{Bethe1990,Logue2012,Andersson2013,Powell2016}). While this makes it difficult to predict the GW signal and its detectability, it represents a unique opportunity to probe the CCSN inner dynamics, inaccessible to EM observations. 
Promising GW signals associated with CCSNe may also originate from the newly-formed compact object soon after birth, in particular if the latter is a ``millisecond'' NS (i.e.~a NS spinning with millisecond period) (e.g., \cite{Cutler2002,Corsi2009,DallOsso2009,Gualtieri2011,DallOsso2018}). The expected event rates in this case depend on the fraction of millisecond NSs that are born in CCSNe (e.g., \cite{Corsi2009,Beniamini2019}). 
During the 2G GW network era, one may expect GW detections of CCSNe events to be limited within maximum distances that vary, depending on models, from tens of kpc up to a few Mpc (e.g., \cite{Abbott2020SN} and refs.~therein; see also \cite{vanPutten2001}). 
3G detectors, with their $\sim\!10$ times larger sensitivity, will lead to a corresponding extension of the expected horizon and open new prospects for discoveries. 

The detection of the GW signal from a CCSN and/or a newly-born millisecond NS along with EM counterparts would represent a breakthrough discovery for NS physics.
The most relevant EM signals expected in association with such events are those temporally coincident or nearly coincident with the GW burst epoch, since their detection can mark with more precision the start time of the GW emission and this would prove extremely helpful, if not crucial, for the challenging signal search process. Such EM signals are primarily high-energy transients and THESEUS will be perfectly suited to catch them. 

In particular, long GRBs are known to be associated with highly energetic CCSNe, and therefore nearby long GRBs may have a detectable GW counterpart. In this case, the full power of THESEUS as a GRB detector can be exploited. Similarly to short GRBs (Section~\ref{SGRB}), nearby events will be observable up to a certain viewing angle with respect to the GRB jet axis, via the prompt and afterglow emission and possibly also via the extended emission and/or an X-ray plateau.
Low Luminosity GRBs (LLGRBs; e.g., \cite{Toma2007,Virgili2009}) and X-ray Flashes (XRFs; e.g., \cite{Sakamoto2005}) populating the nearby Universe, if associated with CCSNe\footnote{See, e.g., \cite{Ciolfi2016} for an alternative interpretation of XRFs.} detectable in GWs, are also very promising as they are expected to be more numerous than ordinary long GRBs and their softer emission makes them ideal targets for THESEUS. 
In addition, for those CCSNe giving birth to highly magnetized NSs, SDPTs observable by THESEUS/SXI may be produced (as for the highly magnetized NS remnants resulting from NS-NS mergers; see Section~\ref{SDPT}). 
Finally, shock breakout signals associated with SNIbc and SNII explosions are expected to  follow closely the core-collapse (within $\sim\!10\!-\!1000$\,s), appearing as bright X-ray bursts lasting for seconds to tens of minutes and having luminosities in the range $10^{43}-10^{46}$\,erg/s (e.g., \cite{Li2007}). THESEUS/SXI and XGIS can detect such shock breakout signals up to about 50\,Mpc, leading to an estimated rate of the order of one event per year.

Another particular class of GRBs potentially associated with CCSNe and their GW emission are the so called ``ultra-long GRBs'', having a prompt emission lasting for tens of minutes up to several hours (e.g., \cite{Levan2014}). So far, only a small fraction ($\sim\!1\%$) of GRBs have been identified as ultra-long GRBs, which could be due to an intrinsic low rate but also to their lower luminosity (see also \cite{Zhang2014}). A larger accreting mass with respect to ordinary long GRBs has been invoked to explain the exceptional durations, suggesting blue supergiants as well as Pop III stars as possible progenitors (e.g., \cite{Gendre2013}). Another possible explanation is the long-lasting energy injection from a newly-born rapidly spinning NS. 
Also in this case, if a GW signal is detected from such systems, the combination with EM observations will represent a unique opportunity to identify the correct physical scenario. 
On the basis of ultra-long GRB average properties (see, e.g., \cite{Gendre2019}), simulation results show that THESEUS/XGIS will be able to detect these transients up to an average distance of about $z\!\sim\!1$ and with THESEUS/SXI up to very large distances ($z\!\sim\!3$ or more). At the expected (much smaller) distances for a joint GW detection, THESEUS will thus be able to catch ultra-long GRBs even for rather large viewing angles.
\begin{figure}[!t]
  \centering
  \includegraphics[width=1.0\linewidth]{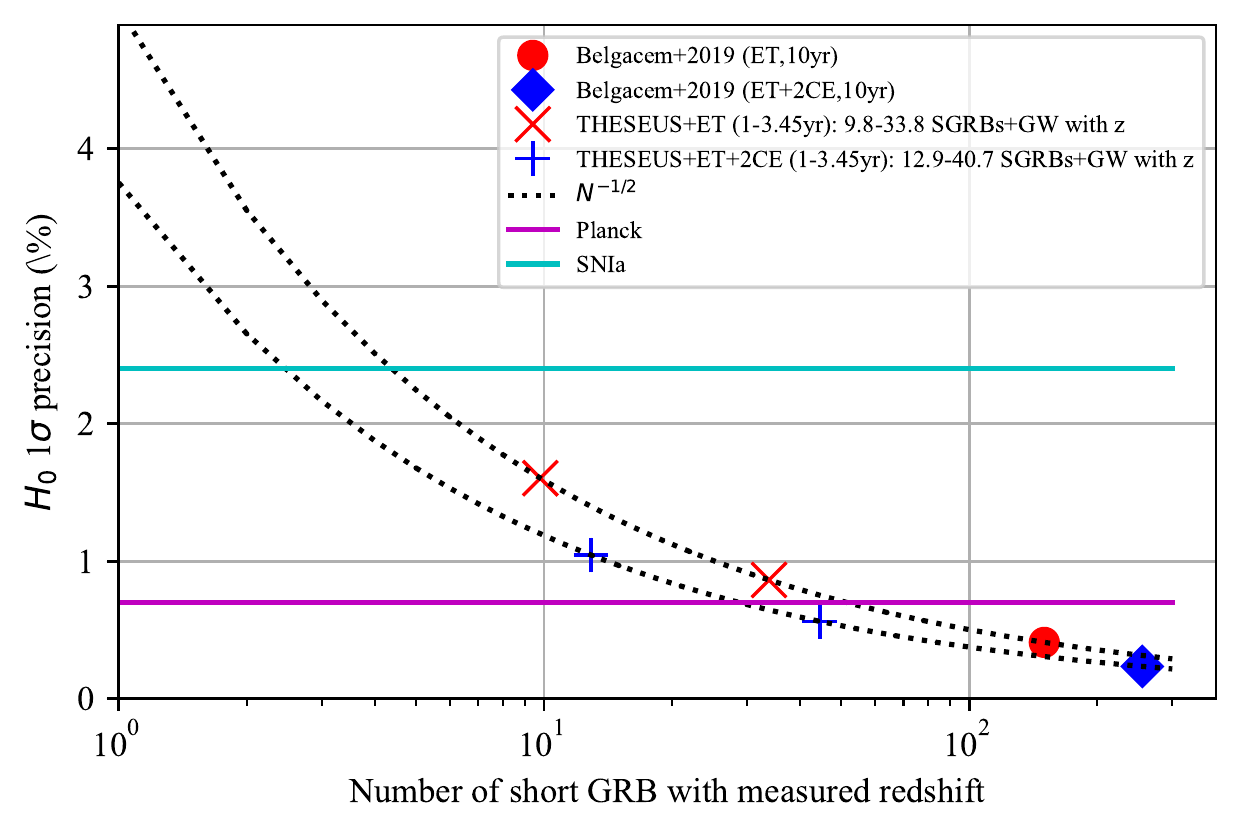}
  \caption{Hubble constant 1 sigma precision level as a function of the number of compact binary mergers for which the redshift and the luminosity distance can be measured from the electromagnetic and gravitational wave emission, respectively. In this plot, the electromagnetic counterpart comes from aligned and misaligned short GRBs (see Table~\ref{TAB_2-3}) for which redshift can be measured. These numbers are conservative estimates from afterglow simulations performed with ``afterglowpy'' (see text).}
   \label{H0_plot}
\end{figure}

\subsection{Magnetars}
\label{magnetars}

A different source of high-frequency GWs can originate from the bursting activity of highly magnetised isolated NSs or magnetars, which are known to manifest themselves as soft gamma repeaters (SGRs) or anomalous X-ray pulsars (AXPs) (e.g., \cite{Mereghetti2008,Turolla2015}).
Such bursting activity is likely associated with dramatic magnetic field readjustments possibly causing fractures of the solid crust on their surface. Of particular interest are the rare ``giant flares'' already observed from three different soft gamma repeaters (e.g., \cite{Thompson1995}; \cite{Mereghetti2008,Turolla2015} and refs.~therein; see also \cite{Svinkin2021}), which inevitably excite strong non-radial oscillation modes that may produce detectable GWs (e.g., \cite{Ioka2001,Corsi2011,Ciolfi2011,Ciolfi2012}). At the typical dominant (i.e.~f-mode) oscillation frequencies in NSs ($\sim$kHz), ET and CE might be sensitive to relatively close giant flare events (see also \cite{LVC-magnetar2019,LVC-long2019,LVC-short2019}).

In terms of EM observations, magnetar bursts are commonly detected in the X-ray and soft gamma-ray bands (e.g., \cite{Mereghetti2008,Turolla2015}).\footnote{Notably, a millisecond-duration radio burst was recently observed from a Galactic magnetar \cite{CHIME2020,Bochenek2020} along with a high-energy counterpart \cite{Mereghetti2020}, strengthening the putative link between magnetar flares and Fast Radio Bursts.} The initial short ($<\!0.5$\,s) bright spikes of giant flares can be detected with THESEUS/XGIS to considerable distances, favoured by its low energy threshold with respect to other coded-mask detectors. The following bursting activity is instead easily detectable with SXI.

\section{Neutrino sources}
\label{neutrino}

High-energy neutrinos provide unique signatures of the presence of accelerated hadrons at the source. Emerging from hadronic collisions with characteristic energies 20 times \\ smaller than the energy of the accelerated protons, the properties of the neutrino events detected so far point towards cosmic objects capable of producing energies as high as EeV. These sources are possibly responsible for the flux of Ultra High Energy Cosmic Rays (UHECRs) as well. Hence, it is of paramount importance to address the question of the origin of the high-energy neutrinos, as they can probe the most extreme accelerators in the Universe. 

Within our Galaxy, no sources are known to date that can achieve EeV energies, except for the cosmic rays possibly interacting with dense proton targets in the Galactic disk. However, these are likely not the major contributors of the diffuse neutrino flux observed \cite{Pagliaroli2016}. On the other hand, many classes of extra-galactic sources are considered plausible candidates, as their energetics can explain neutrino observations (e.g., \cite{Meszaros2017}). Among these, particularly relevant are GRBs, AGNs, and star forming galaxies, which represent targets for THESEUS.
The joint detection of a large number of neutrino and EM emission sources, feasible only during the 2030s with next generation neutrino detectors, will allow us to answer long-standing questions on the acceleration mechanisms inside these systems, on which (hadronic vs.~leptonic) processes characterize the photon and neutrino production, and on the role of different type of sources in producing the observed diffuse neutrino flux. 

While AGNs are thought to produce the largest fraction of such neutrinos, another, smaller fraction of diffuse neutrino emission is expected to originate from SNe in starburst galaxies that are expected to behave as calorimeters \cite{Waxman1999}. 
Starburst galaxies have typically masses $M^*\!< \!10^{9}\,M_{\odot}$ \cite{Brinchmann2004} and obey the relation between $M^*$ and the galaxy K-band luminosity ($L_K$) that is $\log[(M^*/M_{\odot}) (L_{\odot}/L_K)]\!<\!-0.3$ \cite{Arnouts2007,Ilbert2010}, which is valid also for low-mass star-forming galaxies. This implies an absolute magnitude $M_K\!>\!-21.3$. With a H-band limit $\sim\!21$ (AB) and assuming a negligible H-K color, THESEUS/IRT could observe such galaxies up to $z\sim0.6$.

An additional, still very uncertain fraction of neutrino diffuse emission can originate from GRBs. If during the GRB prompt phase a non-negligible fraction of baryons is accelerated at internal shocks \cite{Ghisellini2020}, neutrinos are likely to be produced in proton-photon interactions, given the intense radiation field of the jet. 
So far, no neutrino event has been detected in correlation with a GRB \cite{Aartsen2016,Adrian2013}, indicating a limited neutrino production in the most powerful sources \cite{Albert2017} and strengthening the case for extending this investigation to fainter sources. For this reason, LLGRBs may be better candidates than bright GRBs to account for the IceCube diffuse neutrino flux, although likely not dominant \cite{Murase2006,Denton2018}. The sensitivity and extended energy bandpass of THESEUS are fundamental to probe the poorly-sampled fraction of intrinsically soft LLGRBs (see also Section~\ref{CCSNe}). 
In short GRBs, proton-proton collisions in the post-merger accretion disk are also expected to take place and contribute to the neutrino emission. 
As for long GRBs, no neutrino event has been detected so far in coincidence with a short GRB \cite{Aartsen2016,Adrian2013,Albert2017}. Recent studies have suggested that high-energy neutrinos can be efficiently produced during the Extended Emission phase of short GRBs \cite{Aartsen2016,Adrian2013}, a target that THESEUS/XGIS will detect up to large distances (see also Section~\ref{EE}).

\section{External triggers}
\label{external}

In addition to the major contribution of THESEUS in multi-messenger astronomy in standard survey mode, enabled by its capability to cover large portions of the sky and independently discover the EM counterparts of neutrino or gravitational wave sources, it will also be possible to activate Target of Opportunity (ToO) observations pointing in the direction of a given GW or neutrino trigger event. 
Since the localization of GW and neutrino events can be of the order of a few square degrees or even worse, ToOs with THESEUS will also exploit the large sky coverage of XGIS and SXI to identify the EM counterparts. At the same time, THESEUS capabilities to localize these sources down to a few arcmin will be fundamental to activate further observations via dedicated follow-up campaigns with optical and radio facilities, ultimately characterizing the source and possibly identifying the host galaxy.

While the nominal mission requirement for THESEUS corresponds to a pointing time within about 12 hours since the alert from neutrino or GW detectors, a realistic goal is to follow-up within 4 hours. With such premises, there are a number of potential target signals.
In the context of NS-NS/NS-BH mergers, late time X-ray emission that could be observed by THESEUS a few hours after the merger (i.e.~the GW trigger) is predicted in various forms, including HLE from a structured short GRB jet and, for NS-NS mergers only, SPDT from a long-lived highly magnetized NS remnant (see Sections~\ref{plateau} and \ref{SDPT}). Moreover, short GRB jet afterglows in both X-rays and NIR might have peak times significantly delayed with respect to the initial burst (Section~\ref{afterglow}) and thus be observable in ToO mode by SXI and IRT. Finally, thermal kilonova NIR transients are expected to peak on timescales of days to weeks (Section~\ref{kilonova}) and could therefore be observed with IRT provided that a good localization (of the order of arcminutes) is previously obtained via an optical/IR detection. 
Observations of CCSNe triggered by a GW precursor represent another possible ToO application, aimed at catching, e.g., shock-breakout X-ray signals (in events like SN\,2008D the time-delay can be of several hours).

THESEUS ToO observations of neutrino events will be crucial to enhance the confidence in establishing their cosmic origin and to provide a complete phenomenological picture of the corresponding sources and underlying neutrino production mechanisms. 
Compatible with the THESEUS reaction timescales are, for instance, the flaring activity of AGNs with time-scales of hours/days or NIR observations of star-bursting galaxies within neutrino sky localization region (for well localized events only, i.e.~$<\!1$\,deg$^2$) (see also Section~\ref{neutrino}).

\section{Conclusions}
\label{concl}

Multi-messenger observations of GW and neutrino sources have led to a number of breakthrough discoveries in the last few years. The cases of the short GRB\,170817A and the blazar TXS0506+056 proved that among the most promising EM counterparts of these sources, X-ray and gamma-ray transient signals play a central role. Therefore, high-energy transient sky surveyors will certainly be of the utmost importance for the future of multi-messenger astrophysics. 

Thanks to its unique capabilities, THESEUS will independently detect and characterize the main EM counterparts of multi-messenger sources in an era in which next generation neutrino and GW facilities will ensure much higher detection rates than today. Events like short GRBs, long GRBs and low-luminosity GRBs, AGNs and blazars, as well as X-ray emission from bursting (e.g., SGRs) or spinning-down NSs all represent ideal targets for THESEUS. Moreover, this mission will disseminate alerts of newly discovered multi-messenger sources with accurate sky localisation, which will be crucial to allow coeval narrow field facilities to perform deep follow-up observations. 

Given the design sensitivities of next generation GW detectors (such as ET and CE) and neutrino detectors (such as IceCube-Gen2 and KM3NeT), as well as those of several future radio/optical/X-ray/gamma-ray facilities such as SKA, ELT, CTA, and Athena (among others), the 2030s will be a golden era of multi-messenger astrophysics. The expected launch epoch of THESEUS (early 2030s) and its performances make this mission timely and perfectly suited to face the future challenges posed by the multi-messenger investigation of the transient Universe, offering excellent prospects for a major contribution in the field.

\begin{acknowledgements}
We acknowledge support from the ASI-INAF \\ agreement n.~2018-29-HH.0.
MB and RH acknowledge the support of the European Union's Horizon 2020 Programme under the AHEAD2020 Project grant agreement 871158.
Part of this research was conducted by the Australian Research Council Centre of Excellence for Gravitational Wave Discovery (OzGrav), through project number CE170100004. EH also acknowledges support from an Australian Research Council DECRA Fellowship (DE170100891).
AR acknowledges support from the project Supporto Arizona \& Italia.
LSal acknowledges support from the Irish Research Council under grant  GOIPG/2017/1525.
LH acknowledges support from Science Foundation Ireland under grant \\ 19/FFP/6777 and from the EU AHEAD2020 project (grant agreement 871158). 
PDA acknowledges funding from the Italian Space Agency, contract ASI/INAF n. I/004/11/4. 
MB, PDA, and EP acknowledge support from PRIN-MIUR 2017 (grant 20179ZF5KS). 
MDP acknowledges support for this work by the Scientific and Technological Research Council of Turkey (T\"UBITAK), Grant No: MFAG-119F073.
\end{acknowledgements}

\bibliographystyle{spphys}       
\bibliography{refs}

\end{document}